\documentclass[5p,twocolumn]{revtex4-2} 
\usepackage{graphicx}
\usepackage[table]{xcolor}
\usepackage{epstopdf}

\begin{document}

\oddsidemargin = -12pt
\title{$^{12}$C and $\alpha$-clusters, $0^+$ spectrum, and Hoyle-state candidates in $^{24}$Mg }


\author{J. Cseh}
 \email{cseh@atomki.hu}
\author{G. Riczu}%
 \email{riczugabor@atomki.hu}
 \affiliation{%
 Institute for Nuclear Research, 4001 Debrecen, P. O. Box 51, Hungary\\
}%

 \author{D. G. Jenkins}
 \email{david.jenkins@york.ac.uk}
\affiliation{%
 School of Physics, Engineering and Technology, University of York, Heslington, York, YO10 5DD, United Kingdom\\
}%


\begin{abstract}
\noindent
\textbf{Background:} 
A recent inelastic alpha-scattering experiment \cite{prl22} found $0^+$ resonances in $^{24}$Mg on and above the $^{12}$C+$^{12}$C break-up threshold.  It has been conjectured that the states have a $^{12}$C+$^{12}$C cluster structure, and play a similar role in accelerating $^{12}$C+$^{12}$C fusion to the manner in which the Hoyle state accelerates production of  $^{12}$C in massive stars. 
\\
\textbf{Pupose:} We wish to build up a quantitative theoretical basis for the considerations of the Hoyle-state paradigm, by calculating the distribution of the $0^+$ states in the shell, as well as in the relevant cluster models.
\\
\textbf{Methods:} We determine the spectrum of excited $0^+$ states in $^{24}$Mg nucleus using multiconfigurational dynamical symmetry calculations leading to a unified description of the quartet (or shell), $^{12}$C+$^{12}$C and $^{20}$Ne+$^{4}$He cluster configurations. 
\\
\textbf{Results:} The density of $0^+$ states in the quartet spectrum is comparable to that found in experiment; however, the density of cluster states is considerably less.
\\ 
\textbf{Conclusions:} The recently observed alpha-scattering resonances do not seem to be simple $^{12}$C+$^{12}$C cluster states, but are more plausibly interpreted as fragmented cluster states due to coupling to quartet excitations, as background states.
\end{abstract}

\maketitle

\section{Introduction}

In the 1950s, Fred Hoyle predicted the existence of a resonance in $^{12}$C with spin-parity $0^+$ lying just above the threshold for breakup into three alpha particles, that could explain the production of $^{12}$C in masive stars~\cite{hoyle}. This state, largely responsbile for the existence of carbon in our universe, was subsequently found in experiment and is traditionally known as the ``Hoyle state'' in Hoyle's honour. In a recent work \cite{prl22}, this Hoyle-state paradigm was extended to the $^{12}$C+$^{12}$C fusion. In particular the authors investigated if there are $0^+$ resonance(s) in $^{24}$Mg close to the $^{12}$C+$^{12}$C threshold energy, which can play a similar role to that of the Hoyle state in the carbon burning. Four such $0^+$ resonances were identified in an 
$^{24}$Mg$(\alpha, \alpha^{\prime})$$^{24}$Mg experiment \cite{prl22}; it is argued that some or all of these resonances have
$^{12}$C+$^{12}$C cluster structure, therefore, the phenomenon is parallel with that of the Hoyle-state. The fact that these states predominantly decay by $\alpha$-particle emission is considered to be a further argument in favor of this interpretation.

Inspired by this exciting conjecture and experimental finding, we have carried out calculations in order to build up some quantitative background from the theoretical side. In particular, we have determined the distribution of $0^+$ states in $^{24}$Mg, and we have studied their relation to the $^{12}$C+$^{12}$C  and $^{20}$Ne+$^{4}$He cluster configurations. 
We did so by reproducing the gross features of the energy spectrum over a wide energy range. 

As a theoretical tool, we have applied multiconfigurational dynamical symmetry (MUSY) \cite{musy3,musysym}. This symmetry connects the shell, collective and cluster models to each other in order to handle multi-shell problems; therefore, it is able to give a unified treatment of different phenomena, which are usually described by separate models. This symmetry has turned out to be successful in relating diverse segments of spectra at different regions of excitation energy and deformation, which are produced in various reactions (and have connections to various cluster configurations) \cite{musy3,musysym,musyapp1,musyapp2,musyapp3}.

Previously, the semimicroscopic algebraic cluster model was applied to the description of the $^{24}$Mg spectrum in
\cite{cc93}. That early calculation and the present work have considerable similarities. In both cases, low-energy states and the high-lying 
$^{12}$C+$^{12}$C spectrum of resonances are described in a unified way: by applying semimicroscopic approaches, with dynamically symmetric Hamiltonians.  There are, however, important differences as well. In \cite{cc93}, only the $^{12}$C+$^{12}$C cluster configuration was taken into account, while, here, we consider the quartet states of  $^{24}$Mg, which form the spin-isospin zero sector of the no-core shell model space, together with the $^{12}$C+$^{12}$C and $^{20}$Ne+$^{4}$He clusterizations. For a unified treatment of the spectra of different configurations we apply a Hamiltonian invariant with respect to the transformations from one configuration to the other \cite{musy3,musysym,logic}. According to the semimicroscopic nature of our approach, the energy functional contains free parameters, but only in a very small number and with well-defined physical content, as is discussed in more detail below. The number of fitted parameters of our Hamiltonian is only 3+2 (for the energy and moment of inertia functionals), as opposed to the 12 employed in the previous description in \cite{cc93}. 

The MUSY Hamiltonian provides us with a spectrum of the quartet (or shell) and the two cluster configurations over a wide range of excitation energy. Along the deformation parameter the states cover the moderate ground-region, as well as the largely deformed (theoretically predicted) shape isomers up to the linear alpha-chain. The density of $^{12}$C+$^{12}$C resonances is obtained as a parameter-free prediction, and its comparison with experimental observation serves as a consistency check of the description.
 
In what follows, we first give a brief introduction to MUSY, then apply it to the $^{24}$Mg nucleus with special attention to the relevant cluster configurations. Finally, some conclusions are drawn.

\section{Multiconfigurational dynamical symmetry}

MUSY connects the fundamental structure models of nuclei: the shell model, the quadrupole collective model and the cluster model.  
It is a multi-shell generalization of the historical relationship between these models, established in 1958 for a single shell problem
\cite{elliott1, elliott2, wildkan, baybohr}. In particular, MUSY provides us with a unified multiplet structure of different configurations, as well as with operators for the calculation of physical observables. 

For the single-shell problem, it is defined by the group chain
\begin{eqnarray}
U(3) \ \ \supset \ \ &SU(3)& \ \ \supset \ \ SO(3) \  \\
\nonumber
|[n_1,n_2,n_3], \ \ \ &(\lambda, \mu),& \ \ K,\ \ \ \ L \ \ \ \rangle .
\label{u3}
\end{eqnarray}
The representation labels are the quantum numbers which characterize the basis states of the shell model. The collective and cluster bands are specified by their U(3) symmetries, i.e. they are picked up from the shell model basis. The spin and isospin degrees of freedom are described by the Wigner U$^{ST}$(4) group, and in the case of a U$^{ST}$(4) scalar (spin and isospin zero) representation, the shell model simplifies to a quartet model \cite{quartet}. When a Hamiltonian with a U(3) dynamical symmetry is applied
\begin{eqnarray}
H &=& H_{HO} + \chi QQ + e^{\prime} LL 
\nonumber \\
H &=& C^{(1)}_{U3}+ \chi C^{(2)}_{SU3} + e C^{(2)}_{SO3} ,
\end{eqnarray}
 the energy can be calculated analytically and in a unified way for different configurations. Here, $C$ stands for the invariant operator of the group indicated as a subscript (and of the order shown as a superscript). A similar procedure is applicable for electromagnetic transitions.

For the multi-shell problem, the dynamical symmetry is defined by the group chain
\begin{eqnarray}
U_s(3) \ \otimes \ U_e(3) \supset &U(3)& \supset SU(3) \supset SO(3) \hspace{14pt}  \\
\nonumber
|[n_1^s,n_2^s,n_3^s], [n_1^e,n_2^e,n_3^e], \rho, &[n_1,n_2,n_3]& \ \ \ (\lambda, \mu), \ K,\ \ L \ \ \ \rangle .
\label{u3u3}
\end{eqnarray}
This symmetry is the common intersection of the symplectic shell model
\cite{sympl1,sympl2},
the contracted symplectic model,
\cite{contrsympl1,contrsympl2},
and the semimicroscopic algebraic cluster model
\cite{sacm1,sacm2}.
These models are many-major-shell algebraic approaches, based on the shell, quadrupole and cluster pictures, with microscopic model spaces.
$U_s(3)$ is the symmetry of the lowest-lying shell in the shell and collective models, while in the cluster model, it is the intrinsic symmetry of the clusters. 
$U_e(3)$ gives the major shell excitations, which take place in steps of $2 \hbar \omega$ in the symplectic and contracted symplectic models, and in steps of $1 \hbar \omega$ in the cluster model.

When dynamically symmetric Hamiltonians are applied, which are expressed in terms of invariant operators of the united groups of chain 
(\ref{u3}), 
the operators are invariant with respect to transformations from one configuration to the other
\cite{musy3, logic};
therefore, the energy spectra of states (with common U(3) symmetry) in different configurations are identical.
Here we use the following Hamiltonian:
\begin{equation}
H = \hbar \omega n + a C^{(2)}_{SU3} + b C^{(3)}_{SU3} + d\hat{K}^2 + 
{1 \over {2 \theta}} C^{(2)}_{SO3} .
\end{equation} 
The moment of inertia $\theta$ can be calculated either for ellipsoids (with cylindrical symmetry) defined by the U(3) quantum numbers,
 or it can be expressed in terms of the  $ \theta = a_0 +a_1 C^{(2)}_{SU3}$  invariant operator
 \cite{theta}.


\section{The spectrum of the $^{24}$Mg nucleus}


\subsection{Background}
The experimental data of the low-energy region are compiled in \cite{nndc}. Two bands are well-established from the experimental side, and the energy and spin-parity of many states are known. Some spin-parities are uncertain. They are shown in Figures 1 and 2.

\begin{figure}[h!]
\begin{center}
\includegraphics[height=7.4cm,angle=0.]{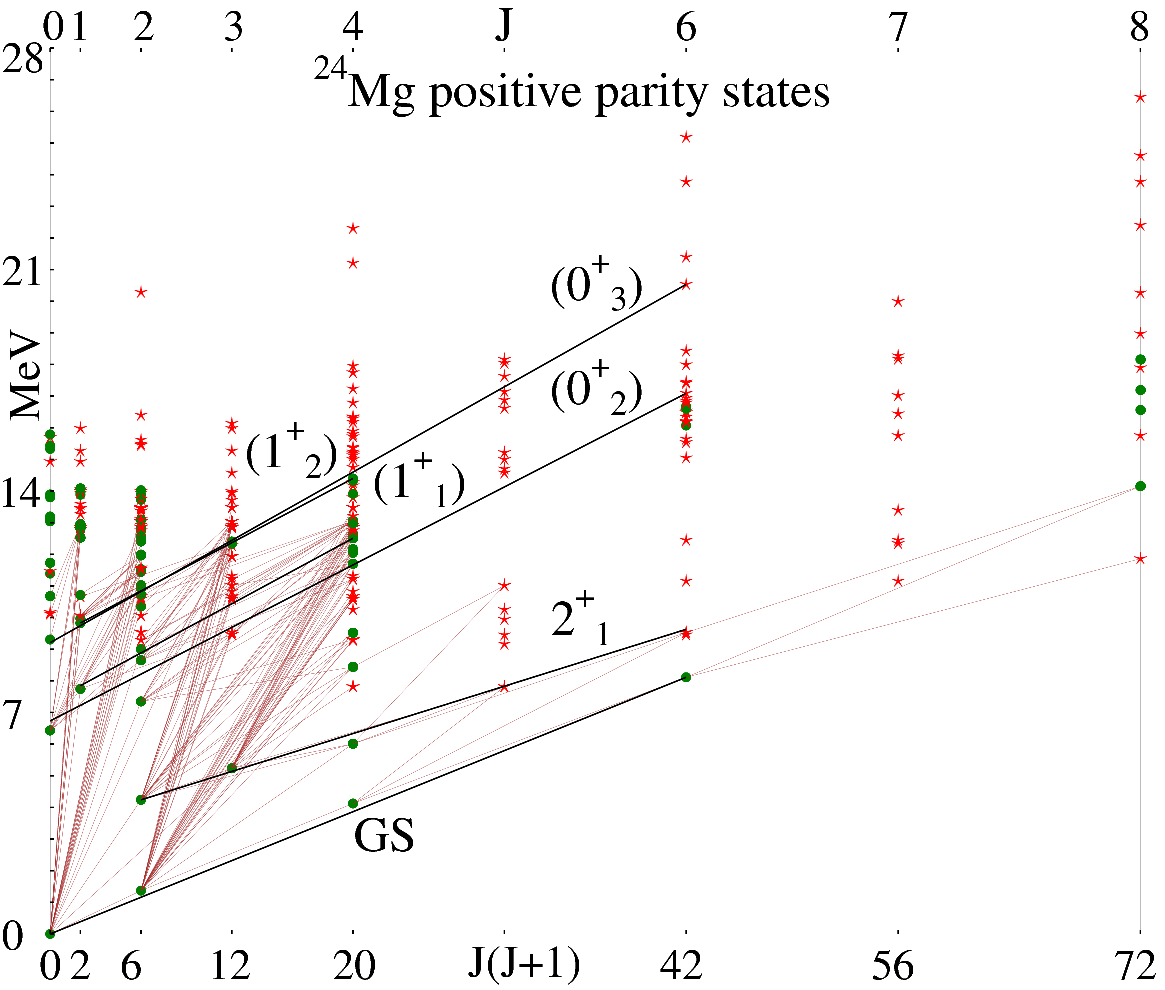}
\caption{Energy of positive parity states as a function of J(J+1) in the $^{24}$Mg nucleus. The experimental data are from ref. \cite{nndc}. Green dots indicate definite spin-parity, while red stars show the uncertain ones. Red lines between states indicate gamma transitions. The black lines show the assumed rotation bands, which we have labelled with K$^{\pi}$ quantum numbers. Only the GS and 2$^+_1$ bands are available in the experimental database (\cite{nndc}).}
\end{center}
\end{figure}

\begin{figure}[h!]
\begin{center}
\includegraphics[height=7.4cm,angle=0.]{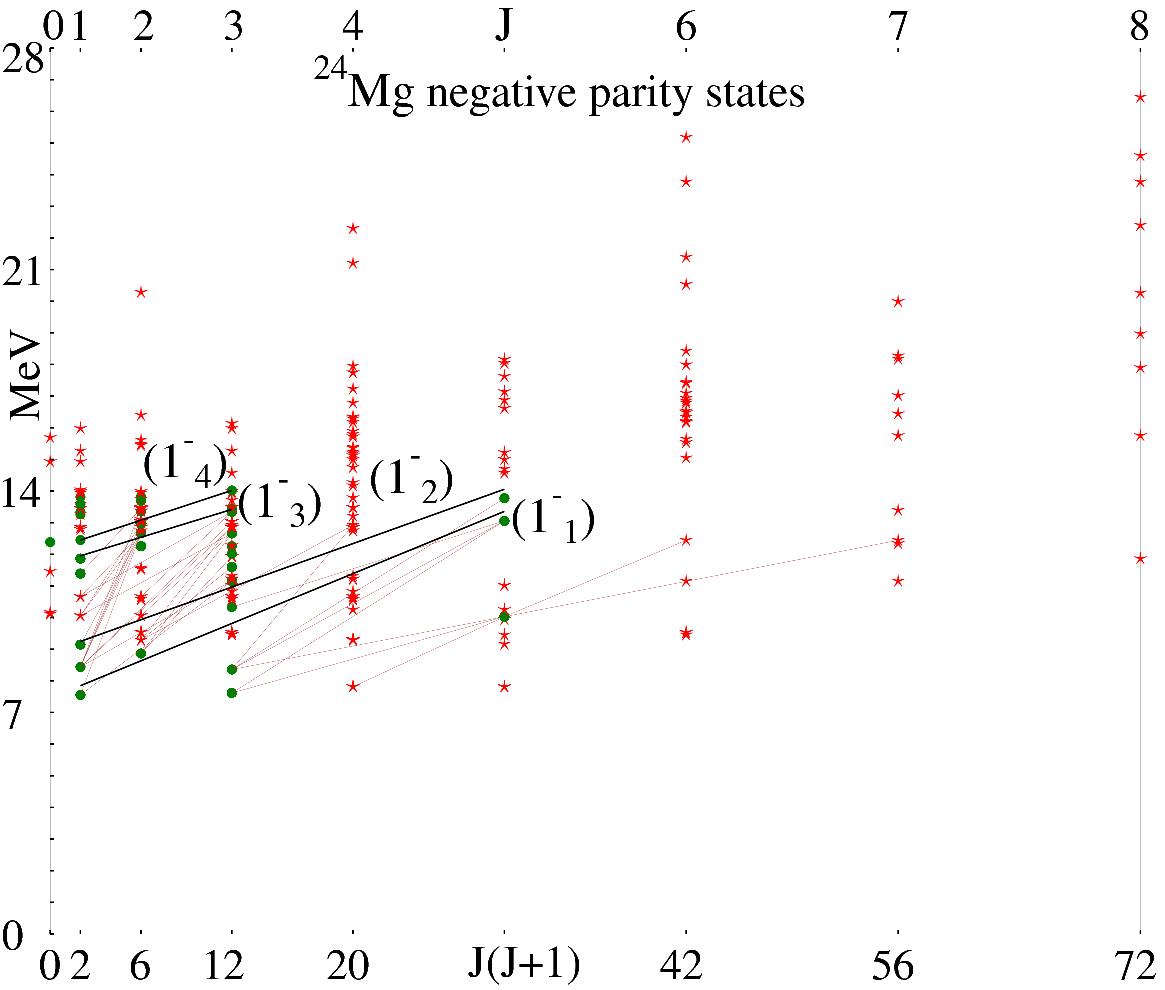}
\caption{Energy of negative parity states as a function of J(J+1) in the $^{24}$Mg. The experimental data are from ref. \cite{nndc}. The notations are the same as in Figure 1.}
\end{center}
\end{figure}

Data on high-lying $^{12}$C+$^{12}$C resonances are collected in \cite{Abbondanno}. Theoretical predictions for shape isomers are available from energy-surface calculations based on the Nilsson model
\cite{larsleand}, 
and from the Bloch-Brink alpha-cluster model
\cite{bb1, bb2, bb3}.
The stable shapes were also determined from the stability and self-consistency calculation of the SU(3) symmetry (or quadrupole deformation)
\cite{lepcsok}, and the result was compared in \cite{shape} with those of energy-minima considerations.


\subsection{Model spaces}
We construct the U$^{ST}$(4)-scalar  (i.e. spin-isospin zero) sector of the no-core shell model, which we call quartet-model space
\cite{quartet}. 
When doing so, the Pauli-exclusion principle is fully observed, and spurious excitations of the center of mass are excluded. 

Cluster model spaces are obtained by characterizing the intrinsic structure of the clusters by their leading U(3) representations, i.e.
[4,4,0] for $^{12}$C,
[0,0,0] for $^{4}$He, and
[12,4,4] for $^{20}$Ne,
and coupling it with a large variety of their relative motion
\cite{sacm1,sacm2}. 
Physically, it means that not only the ground state, but also the rotational band built on the ground state are included in the description as the internal cluster degree of freedom, thus any relative orientations of the clusters and the molecular axes are allowed. In the case of
$^{12}$C+$^{12}$C clusterization, a further symmetry requirement has to be taken into account due to the component clusters being identical.
The most relevant representations are listed in Table I, together with their multiplicities both in the shell and in the cluster models. When the multiplicity is one both in the shell and in the cluster configurations, their wave functions are identical with each other due to the fact that in the shell-model expansion of the cluster state only a single term is included in the superposition. (Basis states of different U(3) representations are orthogonal to each other.)

\begin{table}[!h]
\centering
\resizebox{8.7 cm}{!}
{
\begin{tabular}{|c|c|c|c|c|c|c|}
\hline
n &  U(3) & SU(3) & C$^2$ & quartet & $^{12}C+^{12}C$ & $^{20}Ne+\alpha$ \\
\hline
	& [16, 8,4] & (8, 4) & 148 & 1 & 1 & 1 \\
\cline{2-7}
	& [15, 8, 5] & (7, 3) & 109 & 1 & 1 & \\
\cline{2-7}
	& [14,10,4] & (4, 6) & 106 & 1 & & 1 \\
\cline{2-7}
	& [15, 7, 6] & (8, 1) & 100 & 1 & 1 & \\
\cline{2-7}
0	& [14, 9, 5] & (5, 4) & 88 & 1 & & \\
\cline{2-7}
	& [12, 12, 4] & (0, 8) & 88 & 1 & & 1 \\
\cline{2-7}
	& [14, 8, 6] & (6, 2) & 76 & 2 & 2 & \\
\cline{2-7}
	& [13, 10, 5] & (3, 5) & 73 & 1 & & \\
\cline{2-7}
	& [13, 9, 6] & (4, 3) & 58 & 1 & & \\
\cline{2-7}
	& [13, 8, 7] & (5, 1) & 49 & 2 & 1 & \\
\hline
\hline
	& [18, 7, 4] & (11, 3) & 205 & 1 & 1 & 1 \\
\cline{2-7}
	& [17, 8, 4] & (9, 4) & 172 & 2 & 2 & 1 \\
\cline{2-7}
	& [16, 10, 3] & (6, 7) & 166 & 1 & & \\
\cline{2-7}
	& [17, 7, 5] & (10, 2) & 160 & 3 & 2 & \\
\cline{2-7}
1	& [17, 6, 6] & (11, 0) & 154 & 1 & & \\
\cline{2-7}
	& [16, 9, 4] & (7, 5) & 145 & 4 & & 1 \\
\cline{2-7}
	& [14, 12, 3] & (2, 9) & 136 & 1 & & \\
\cline{2-7}
	& [16, 8, 5] & (8, 3) & 130 & 7 & 2 & \\
\cline{2-7}
	& [15, 10, 4] & (5, 6) & 124 & 6 & & 1 \\
\cline{2-7}
	& [16, 7, 6] & (9, 1) & 121 & 7 & 1 & \\
\hline
\hline
	& [20, 6, 4] & (14, 2)	&	276	&	1	&	1	&	1	\\
\cline{2-7}
	& [19, 8, 3] & (11, 5)	&	249	&	1	&	1	&	\\
\cline{2-7}
	& [19, 7, 4] & (12, 3)	&	234	&	4	&	2	&	1	\\
\cline{2-7}
	& [19, 6, 5] & (13, 1)	&	225	&	3	&	1	&	\\
\cline{2-7}
2	& [18, 9, 3] & (9, 6)		&	216	&	2	&	&	\\
\cline{2-7}
	& [18, 8, 4] & (10, 4)	&	198	&	13	&	3	&	1	\\
\cline{2-7}
	& [16, 12, 2] & (4, 10)	&	198	&	1	&	&	\\
\cline{2-7}
	& [17, 10, 3] & (7, 7)		&	189	&	4	&	&	\\
\cline{2-7}
	& [18, 7, 5] & (11, 2)	&	186	&	15	&	1	&	\\
\cline{2-7}
	& [18, 6, 6] & (12, 0)	&	180	&	10	&	1	&	\\
\hline
\hline
	&	[24,4,4]	&	(20,0)	&	460	&	1	&	1	&	1	\\
\cline{2-7}
	&	[23,6,3]	&	(17,3)	&	409	&	1	&	1	&		\\
\cline{2-7}
	&	[23,5,4]	&	(18,1)	&	400	&	3	&	1	&	1	\\
\cline{2-7}
	&	[22,8,2]	&	(14,6)	&	376	&	1	&	1	&		\\
\cline{2-7}
4	&	[22,7,3]	&	(15,4)	&	358	&	6	&	1	&		\\
\cline{2-7}
	&	[22,6,4]	&	(16,2)	&	346	&	20	&	2	&	1	\\
\cline{2-7}
	&	[22,5,5]	&	(17,0)	&	340	&	8	&		&		\\
\cline{2-7}
	&	[21,9,2]	&	(12,7)	&	334	&	2	&		&		\\
\cline{2-7}
	&	[21,8,3]	&	(13,5)	&	313	&	22	&	2	&		\\
\cline{2-7}
	&	[16,16,0]	&	(0,16)	&	304	&	1	&		&		\\
\hline
\end{tabular}
}
\caption{The 10--10 most deformed representations of the 0, 1, 2 and 4 $\hbar\omega$ (T=0) quartet model space of the $^{24}$Mg nucleus. Here n denotes the major shell excitation, 
C$^2$ gives the expectation value of the second order Casimir invariant of SU(3),
and the last three columns show the multiplicities of the representations in the quartet and cluster model spaces.}
\label{Table 1}
\end{table}

\subsection{Calculation of the spectrum}
MUSY is usually applied with a simple dynamically-symmetric Hamiltonian
\begin{equation}
\hat{H}=\hbar\omega \hat{n}+a\hat{C^2}_{SU(3)}+b\hat{C^3}_{SU(3)}+d\hat{K}^2+\frac{1}{2\theta}\hat{C^2}_{SO(3)},
\end{equation}
where the first term is the harmonic oscillator Hamiltonian, which can be taken from systematics
\cite{hw}, giving $\hbar\omega=12.595$ MeV for $^{24}$Mg. 
The parameters $a,b,d, a_0, a_1$ are fitted to experimental data.
The second term includes the quadrupole-quadrupole interaction, as in the Elliott model
\cite{elliott1}, 
the next term can distinguish between prolate and oblate deformation, and the last one is the rotational part. $\hat{K}^2$ lifts the degeneracy of states with the same angular momentum within an SU(3) multiplet.

In the fitting procedure, we included the members of the two well-established bands (with weight 1.0, except the uncertain 6$^+$ state of 2$^+_1$ band, which has a weight of 0.5.), and took into account four high-lying shape isomers (with weight 0.1), which were obtained from three different model calculations in good agreement with each other
\cite{shape}. They were needed in order to be able to fit our parameters, since the two well-established low-lying bands belong to a single U(3) representation.  Then, we arranged the experimentally observed states into bands, according to the suggestion of the model spectrum. The ones with known spin parity were also weighted by 0.1, while the uncertain ones were weighted with 0.05.
The goodness of the fit is measured by
\begin{equation}
F=\sum_iw(i)\frac{(E^{exp}_i-E^{th}_i)^2}{(E^{exp}_i)^2}.
\end{equation}

The values of the parameters are as follows:
$a$=-0.148228 MeV, 
$b$=0.000392 MeV, 
$d$=0.666544 MeV, 
$a_0$=0.141965${\hbar^2 \over {MeV}}$, 
$a_1$=0.015811${\hbar^2 \over {MeV}}$,
belonging to $F=0.11378$.
Figure 3 shows the model spectrum of these parameters in comparison with the experimental one.

\begin{figure}[h!]
\begin{center}
\includegraphics[height=5.3cm,angle=0.]{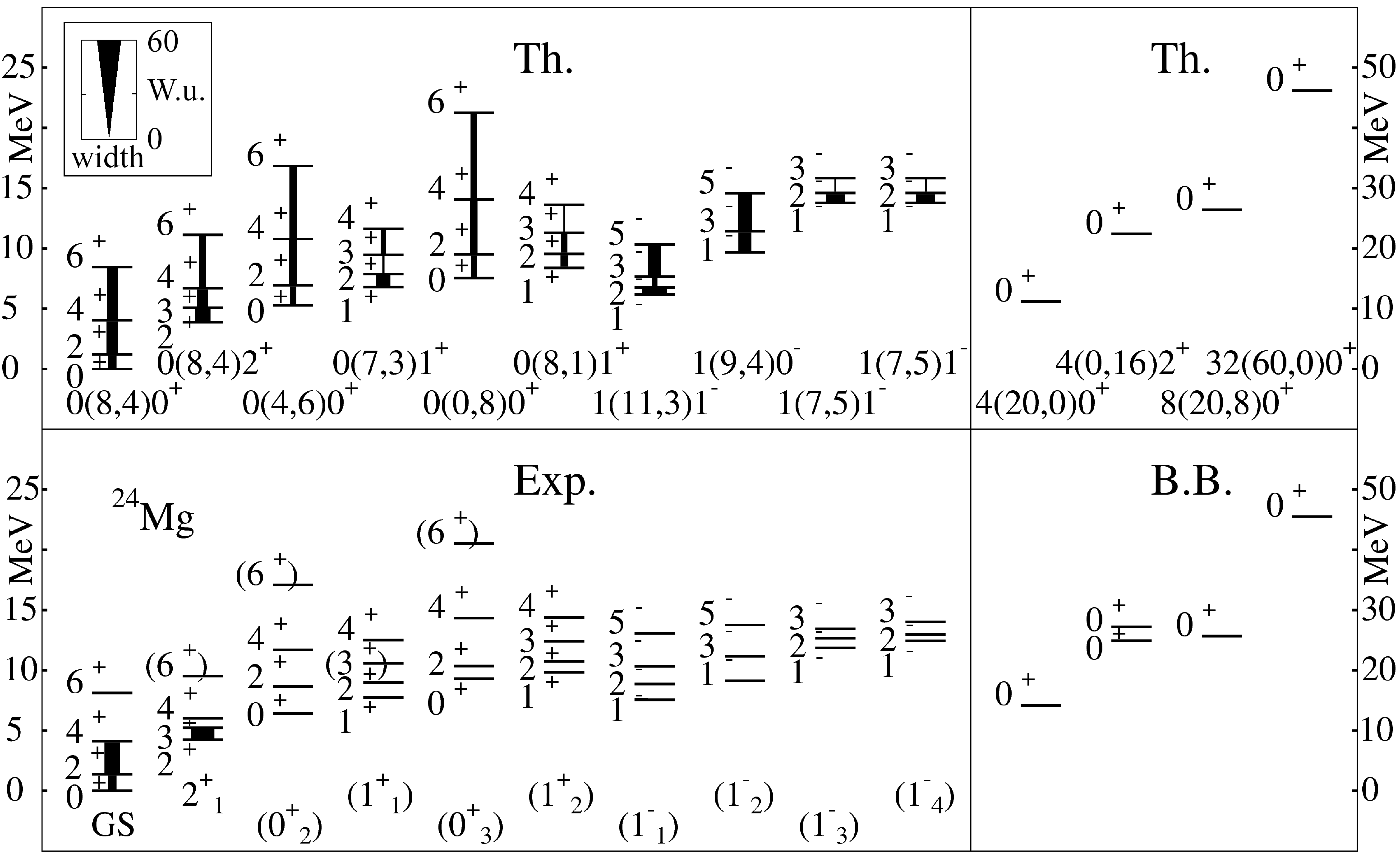}
\caption{The spectrum of the semimicroscopic algebraic quartet model (upper part) in comparison with experimental data for the $^{24}$Mg nucleus (lower left part) and the results of Bloch-Brink $\alpha$ cluster model (lower right part). Experimental bands are labeled by K$^{\pi}$ quantum numbers, and the model states by n($\lambda,\mu$)K$^{\pi}$ labels. The width of the arrows between the states is proportional to the strength of the E2 transition.}
\end{center}
\end{figure}


The in-band E2 transitions are given by the formula \cite{be2}:
\begin{eqnarray}
B(E2,L_i\rightarrow L_f) &=& \frac{|\langle L_f||T^2||L_i\rangle|^2}{2L_i+1}= \\ \nonumber
=\frac{2L_f+1}{2L_i+1}\alpha^2|\langle&(\lambda\mu)&KL_i,(11)2||(\lambda\mu)KL_f\rangle|^2C^2_{SU(3)},
\end{eqnarray}
where the $\alpha^2=0.817$ W.u.
parameter is obtained from the B(E2;$2^+_1 \rightarrow 0^+_1$) transition strength of 21.5 W.u.
\cite{e2}.

\subsection{$^{12}$C+$^{12}$C resonances}
Once the parameters of the Hamiltonian are determined from the quartet spectrum, MUSY determines the cluster spectra free from ambiguity. The predictions for the cluster spectra are shown in Figure 4. These high-lying resonances are not arranged, of course, in rotational bands, therefore, the relevant comparison can be made between the number of states with a definite spin-parity in the energy window, where experiments could be carried out. This is also shown in Table II.

\begin{figure}[h]
\begin{center}
\includegraphics[height=5.3cm,angle=0.]{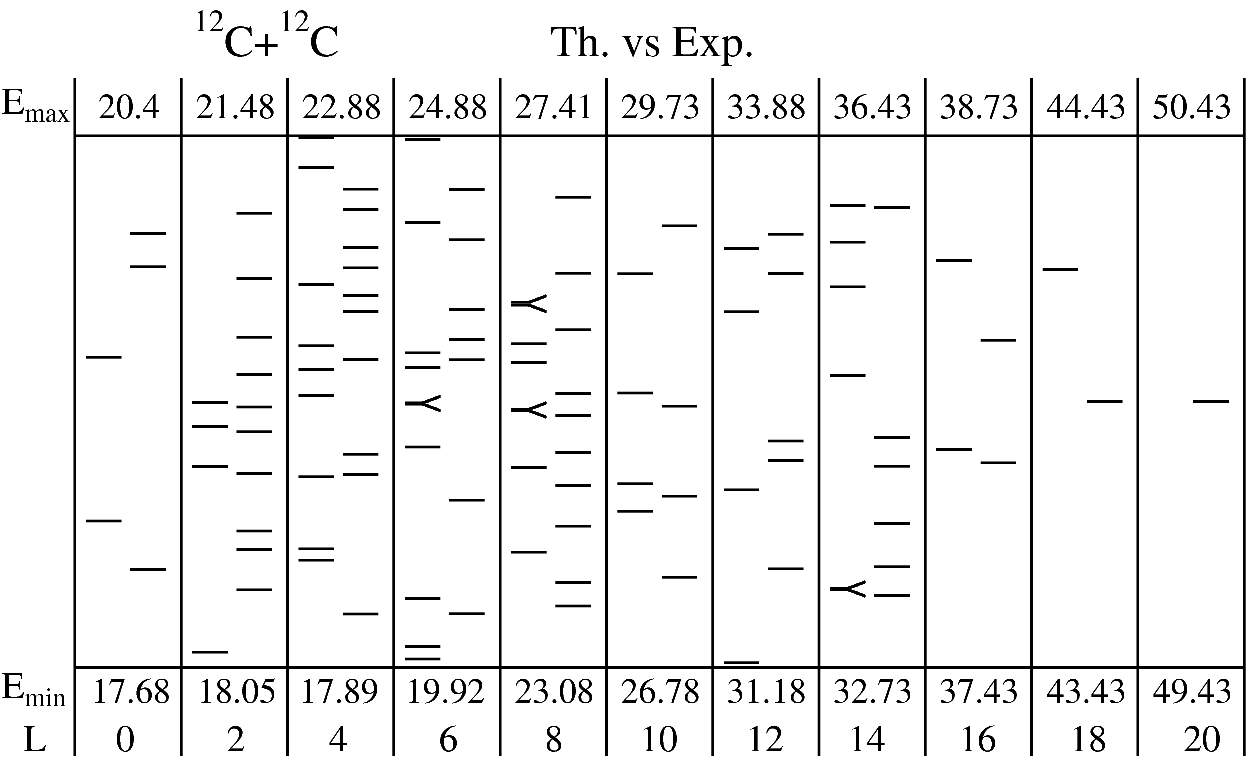}
\caption{Theoretical spectra of even-momentum states of $^{12}$C+$^{12}$C clusterization compared to experimental data in different energy windows. For each angular momentum, the bottom of the energy window is min(E$_{Exp}$)-0.5 MeV and the top is max(E$_{Exp}$)+0.5 MeV. The branchings indicate states that are close to each other. Each column shows the experimental states on the right and the theoretical states on the left.}
\end{center}
\end{figure}

\begin{table}[!h]
\centering
\resizebox{6.0 cm}{!}
{
\begin{tabular}{|c|c|c|c|c|c|c|c|c|c|c|c|}
\hline
L & 0 & 2 & 4 & 6 & 8 & 10 & 12 & 14 & 16 & 18 & 20 \\
\hline
N$_{Exp.}$ & 3 & 10 & 10 & 7 & 10 & 4 & 5 & 6 & 2 & 1 & 1 \\
\hline
N$_{Th.}$ & 2 & 4 & 9 & 10 & 8 & 4 & 4 & 6 & 2 & 1 & 0 \\
\hline
\end{tabular}
}
\caption{Number of $^{12}$C+$^{12}$C resonance states in energy windows indicated by the experimental observations. The energy windows of the theoretical calculation were set between min(E$_{Exp.}$)-0.5  and max(E$_{Exp.}$)+0.5 MeV.}
\label{Table 1}
\end{table}




The large-scale distribution of band-heads of the quartet, $^{12}$C+$^{12}$C, and $^{20}$Ne+$^{4}$He configurations are shown in Figure 5.

\begin{figure}[h]
\begin{center}
\includegraphics[height=11.1cm,angle=0.]{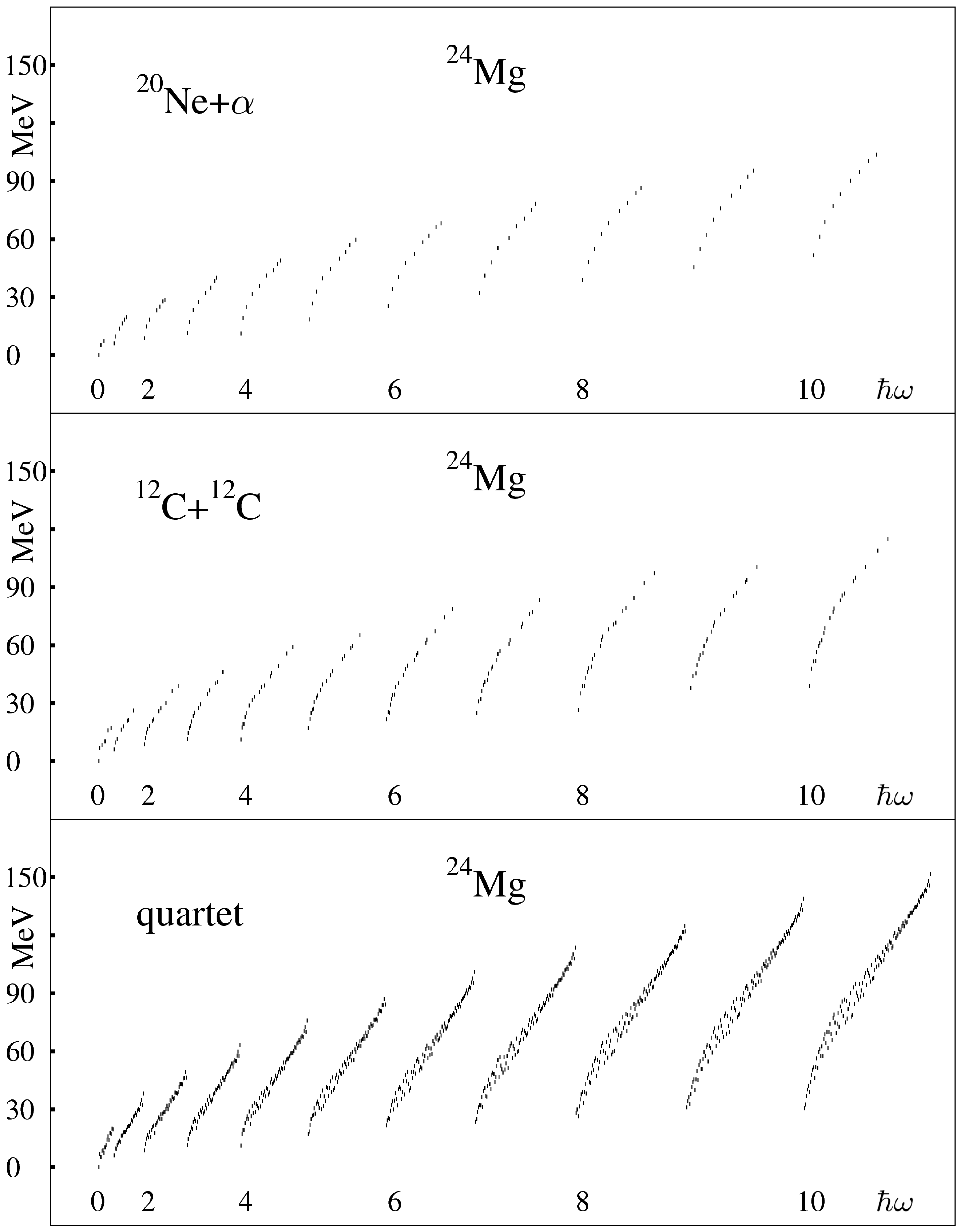}
\caption{The landscape of the quartet and cluster band-heads in the $^{24}$Mg nucleus.}
\end{center}
\end{figure}

\section{The $0^+$ spectra}

The extension of the Hoyle-state paradigm is based on the existence and location of $0^+$ resonances, which have a strong coupling to the $^{12}$C+$^{12}$C and $^{20}$Ne+$\alpha$ reaction channels. Therefore, in this section, we investigate the $0^+$ spectrum of the quartet model, as well as those of the two relevant cluster configurations. These spectra are given in Figure 6, which also locate the 
$^{20}$Ne+$\alpha$ and $^{12}$C+$^{12}$C threshold energies. 

\begin{figure}[h]
\begin{center}
\includegraphics[height=8.5cm,angle=0.]{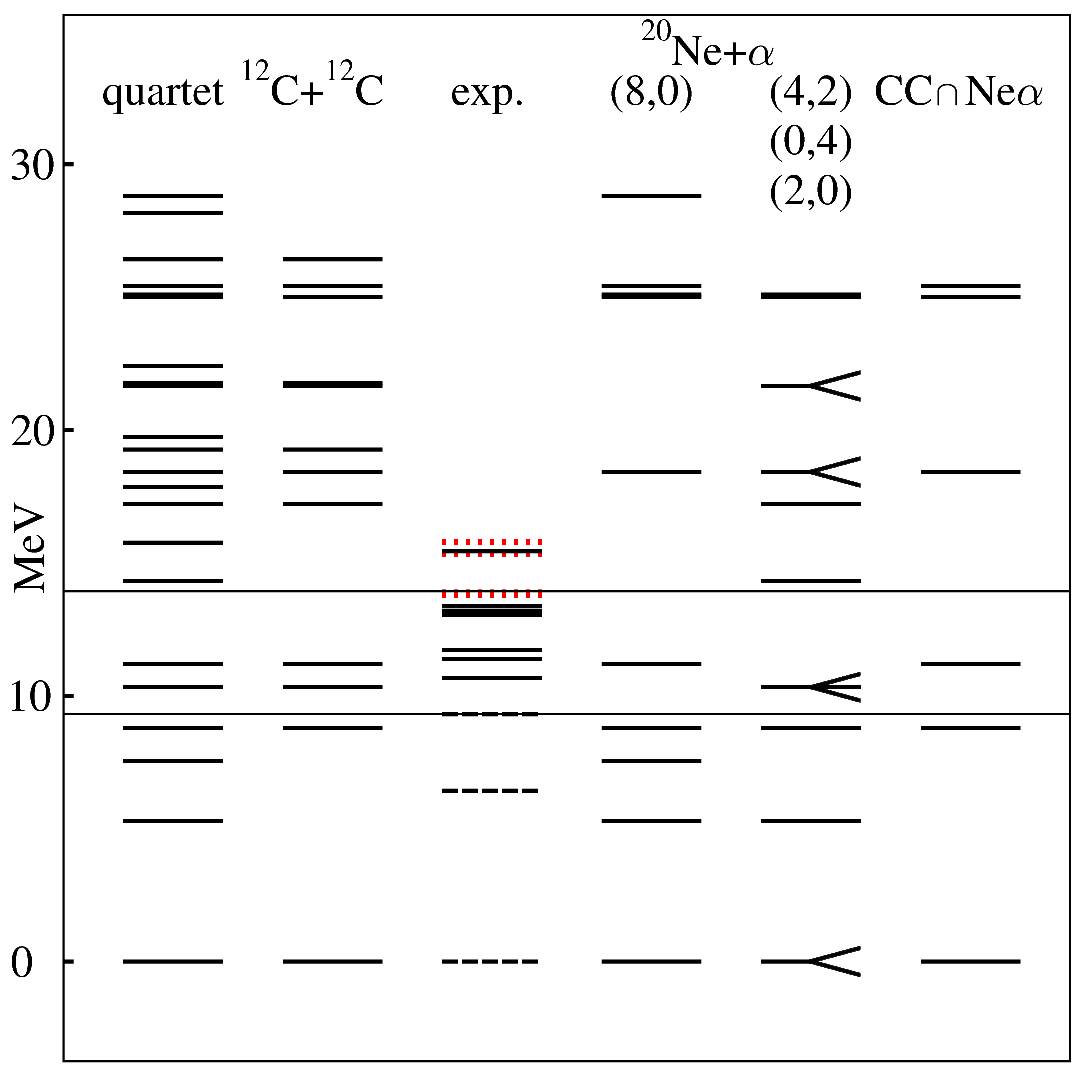}
\caption{0$^+$ spectra in quartet and cluster model spaces in $^{24}$Mg. The lower line represents the $^{20}$Ne+$\alpha$ threshold energy and the upper one the $^{12}$C+$^{12}$C threshold energy. The second from the right spectrum represents the core-plus-alpha configuration with excited  $^{20}$Ne (see text for further explanation.) The branchings denote states with multiple multiplicities. Dashed lines indicate the states also visible in the low-energy spectra, and red dotted lines indicate the states found in \cite{prl22}.}
\end{center}
\end{figure}

As mentioned above, cluster configurations are characterized by dominant U(3) symmetries of the internal structure of the clusters, i.e. the coupling to the reaction channels are handled in terms of the leading representation. As can be seen, the density of $0^+$ resonances of the quartet spectrum is comparable with the observed one, both in general, and in the neighborhood of the $^{12}$C+$^{12}$C threshold energy. The density of such $0^+$ states, however, is smaller for those which overlap with the
$^{12}$C+$^{12}$C and with the $^{20}$Ne+$\alpha$ reaction channels, and especially those which overlap with both. This observation seems to indicate that not all the $0^+$ states observed in the recent experiment
\cite{prl22} have a structure, which contains $^{12}$C+$^{12}$C and $^{20}$Ne+$\alpha$ cluster configurations in the leading representation approximation.

Let us think in terms of the reaction picture, which pays attention to the intermediate region (in between the entrance and exit channels) when all the nucleons are in the close neighborhood of each other within the range of the strong interaction. Then reactions can be classified according to the population of the intermediate states, which are organized into hierarchic order: simple states, doorway states, hallway states... 
\cite{reaction1,reaction2,reaction3,reaction4}. From this viewpoint, the dense pattern of $0^+$ states observed near the threshold energy seems to emerge from the fragmentation of the simple (cluster) state due to the coupling to some background states. Here, we investigate one possibility of this kind from the microscopic viewpoint. 

In the cluster configurations considered here so far, which give the connection to the reaction channels, each cluster is characterized by a  single U(3) representation, i.e. by a single intrinsic state of well-defined quadrupole shape. The intrinsic cluster structure is coupled to the relative motion of the clusters on the U(3) level---the so-called strong coupling cluster scheme.  Here, we consider the possibility of allowing other intrinsic cluster states without major-shell excitation, i.e. we take into account $0 \hbar \omega$ intrinsic cluster excitations. For 
$^{12}$C there is no other U(3) symmetry in the valence shell. For $^{20}$Ne the [12,4,4], [10,6,4], [8,8,4], [8,6,6] U(3) representations can be taken into account. (The sequence of this order corresponds to the decreasing values of the second order SU(3) Casimir invariants, i.e. decreasing quadrupole deformation.) Figure 6 shows the distribution of the $0^+$ states for these larger model spaces too.

\section{Summary and conclusions}
In this paper we have investigated some questions raised by the notion of extending the Hoyle-state paradigm to 
$^{12}$C+$^{12}$C fusion \cite{prl22}. The key concept is that there is (are) $0^+$ resonance(s) near the threshold energy, in common with the triple-$\alpha$ reaction which produces the $^{12}$C nucleus.
In \cite{prl22}, four candidate states were reported from an experiment of inelastic alpha-scattering on $^{24}$Mg.  The preference of these states to break-up into the $^{20}$Ne+$\alpha$ decay channel is considered to be a further support for this interpretation.

Here, we have investigated the $0^+$ spectra of different configurations of $^{24}$Mg. In particular, we have considered the quartet states, i.e. U$^{ST}$(4) Wigner-scalar (spin-isospin zero) sector of the no-core shell model, and the $^{12}$C+$^{12}$C, as well as $^{20}$Ne+$^{4}$He cluster configurations, and their intersection. 

In calculating spectra, we have applied Hamiltonians suggested by the multiconfigurational dynamical symmetry, which is able to handle different configurations in a unified way. Its phenomenological energy functional contains 3+2 fitting parameters for the energy functional and the moment of inertia. In this procedure, the states of the two rotational bands established by the experiments
\cite{nndc}
were taken into account with weight 1.0, while those which we could arrange into bands from the experimental compilation
\cite{nndc},
as well as the theoretically predicted shape isomers of the highly excited and deformed region (from three different models)
\cite{shape}
had weight 0.1. The model spectrum approximates the experimental one reasonably well (Figures 3, 4, 5). 

Based on the description of the low-energy quartet spectrum, MUSY is able to give a parameter-free prediction for the distribution of the states of the high-lying $^{12}$C+$^{12}$C resonances, as shown in Figure 4 and Table II. The similarity of the density of model states to those of the observed ones is considered as a support for the reliability of the MUSY description. 

The distribution of $0^+$ states of different configurations shows the following pattern: the density of the quartet states, both in general, and in the threshold energy region, is similar to that of the experimental one. The number of $0^+$ states in both cluster configurations, and especially in their intersection is considerably less. This shows that the recently observed four $0^+$ resonances are fragmented, and are not simple   
$^{12}$C+$^{12}$C cluster states, with intrinsic ground states of both clusters. 
We have carried out some kind of quantitative comparison in this respect. If one takes into account all the quartet states as background states for the coupling, the densities from the experimental and theoretical sides are comparable. In the 
$^{12}$C+$^{12}$C cluster configuration, major-shell excitations are needed in the internal cluster structure for the observed density of $0^+$ states.

In short, the $0^+$ resonances recently identified in experiment
\cite{prl22}
do not seem to be simple 
$^{12}$C+$^{12}$C cluster states, but their existence is in line with the picture of fragmenting a simple state by coupling it to the $0^+$ quartet states.

{\it Acknowledgements}
This work was supported by the National Research, Development and Innovation Fund of Hungary, financed under the K18 funding scheme with Project No. K 128729.

{}


\section{References}
\begin{thebibliography}{00}
\bibitem{prl22} P. Adsley, M. Heine, D. G. Jenkins, S. Courtin, R. Neveling, J. W. Brümmer, L. M. Donaldson, N. Y. Kheswa, K. C. W. Li, D. J. Marín-Lámbarri, P. Z. Mabika, P. Papka, L. Pellegri, V. Pesudo, B. Rebeiro, F. D. Smit, and W. Yahia-Cherif, Extending the Hoyle-State Paradigm to $^{12}$C+$^{12}$C Fusion, \textcolor{blue}{Phys. Rev. Lett. 129, 102701 (2022).}
\bibitem{hoyle} F. Hoyle, On Nuclear Reactions Occuring in Very Hot STARS.I. the Synthesis of Elements from Carbon to Nickel, \textcolor{blue}{Astrophysic. J. Supp. 1, 121 (1954).}
\bibitem{musy3} J. Cseh, Microscopic structure and mathematical background of the multiconfigurational dynamical symmetry, \textcolor{blue}{Phys. Rev. C 103, 064322 (2021).}
\bibitem{musysym} J. Cseh, G. Riczu, and J. Darai, A symmetry in-between the shapes, shells, and clusters of nuclei, \textcolor{blue}{Symmetry 15, 115 (2023).}
\bibitem{musyapp1}  J. Cseh and G. Riczu, Quartet excitations and cluster spectra in light nuclei, \textcolor{blue}{Phys. Lett. B 757, 312 (2016).}
\bibitem{musyapp2} G. Riczu and J. Cseh, Gross features of the spectrum of the $^{36}$Ar nucleus, \textcolor{blue}{Int. J. Mod. Phys. E 30, 2150034 (2021).}
\bibitem{musyapp3} G. Riczu and J. Cseh, A unified description of spectra of different configurations, deformation and energy regions, \textcolor{blue}{Bulg. J. Phys. 48, 524 (2021).}
\bibitem{cc93} J. Cseh, G. L\'evai, and W. Scheid, Algebraic $^{12}$C+$^{12}$C cluster model of the $^{24}$Mg nucleus, \textcolor{blue}{Phys. Rev.  C 48, 1724 (1993).}
\bibitem{logic} J. Cseh, On the Logical Structure of Composite Symmetries in Atomic Nuclei, \textcolor{blue}{Symmetry 2023 15(2), 371 (2023).}

\bibitem{elliott1} J. P. Elliott, Collective motion in the nuclear shell model. I. Classification schemes for states of mixed configurations, 
\textcolor{blue}{Proc. R. Soc. Lond. A 245, 128 (1958).}
\bibitem{elliott2} J. P. Elliott, Collective Motion in the Nuclear Shell Model. II. The Introduction of IntrinsicWave-Functions, \textcolor{blue}{Proc. R. Soc. Lond. A 245, 562 (1958).}
\bibitem{wildkan} K. Wildermtuh and Th. Kanellopoulos, The “cluster model” of the atomic nuclei, \textcolor{blue}{Nucl. Phys. 7, 150 (1958).}
\bibitem{baybohr} B. F. Bayman and A. Bohr, On the connection between the cluster model and the SU3 coupling scheme for particles in a harmonic oscillator potential, \textcolor{blue}{Nucl. Phys. 9, 596 (1958).}
\bibitem{quartet} J. Cseh, Algebraic models for shell-like quarteting of nucleons, \textcolor{blue}{Phys. Lett. B 743, 213 (2015).}

\bibitem{sympl1} G. Rosensteel and D. J. Rowe, Nuclear Sp(3, R) Model, \textcolor{blue}{Phys. Rev. Lett. 38, 10 (1977).}
\bibitem{sympl2} G. Rosensteel and D. J. Rowe, On the algebraic formulation of collective models III. The symplectic shell model of collective motion, \textcolor{blue}{Ann. Phys. 126, 343 (1980).}
\bibitem{contrsympl1} D. J. Rowe and G. Rosensteel, Rotational bands in the u(3)-boson model, \textcolor{blue}{Phys. Rev. C 25, 3236 (1982).}
\bibitem{contrsympl2} O. Castanos and J. P. Draayer, Contracted symplectic model with ds-shell applications, \textcolor{blue}{Nucl. Phys. A 491, 349 (1989).}


\bibitem{sacm1} J. Cseh, Semimicroscopic algebraic description of nuclear cluster states. Vibron model coupled to the SU(3) shell model, \textcolor{blue}{Phys. Lett. B 281, 173 (1992).}
\bibitem{sacm2} J. Cseh and G. L\'evai, Semimicroscopic Algebraic Cluster Model of Light Nuclei. I. Two-Cluster-Systems with Spin-Isospin-Free Interactions, \textcolor{blue}{Ann. Phys. (N.Y.) 230, 165 (1994).}

\bibitem{theta} J. Cseh, G. Riczu, Moment of inertia and dynamical symmetry, \textcolor{blue}{Symmetry, in press (2023).}



\bibitem{nndc} https://www.nndc.bnl.gov/nudat3/getdataset.jsp?nucleus=\\24Mg\&unc=NDS
\bibitem{Abbondanno} U. Abbondanno, A collection of data on resonances in heavy-ion reactions, \textcolor{blue}{Report No. INFN/BE —91/11, Trieste1991.}
\bibitem{larsleand} G. Leander and S. E. Larsson, Potential-energy surfaces for the doubly even N = Z nuclei, \textcolor{blue}{Nucl. Phys. A 239, 93 (1975).}
\bibitem{bb1} A. Merchant and W. Rae, Systematics of alpha-chain states in 4N-nuclei, \textcolor{blue}{Nucl. Phys. A 549, 431 (1992).}
\bibitem{bb2} J. Zhang and W. Rae, Systematics of 2-dimensional $\alpha$-cluster configurations in 4N nuclei from $^{12}$C to $^{44}$Ti, \textcolor{blue}{Nucl. Phys. A 564, 252 (1993).}
\bibitem{bb3} J. Zhang, W. Rae, and A. Merchant, Systematics of some 3- dimensional $\alpha$-cluster configurations in 4N nuclei from $^{16}$O to $^{44}$Ti, \textcolor{blue}{Nucl. Phys. A 575, 61 (1994).}
\bibitem{lepcsok} J. Cseh, G. Riczu, and J. Darai, Shape isomers of light nuclei from the stability and consistency of the SU(3) symmetry, \textcolor{blue}{Phys. Lett. B 795, 160 (2019).}
\bibitem{shape} P. Dang, G. Riczu, and J. Cseh, Shape isomers of $\alpha$-like nuclei in terms of the multiconfigurational dynamical symmetry, \textcolor{blue}{Phys. Rev. C 107, 044315 (2023).}

\bibitem{hw} J. Blomqvist and A. Molinari, Collective 0$^-$ vibrations in even spherical nuclei with tensor forces, \textcolor{blue}{Nucl. Phys. A 106, 545 (1968).}

\bibitem{be2} Y. Sun, C. L. Wu, K. Bhatt, and M. Guidry, SU(3) symmetry and scissors mode vibrations in nuclei, \textcolor{blue}{Nucl. Phys. A 703, 130 (2002).} 

\bibitem{e2} R.B. Firestone, Nuclear Data Sheets for A = 24, \textcolor{blue}{Nuclear Data Sheets 108, 2319 (2007).}

\bibitem{reaction1} 
H. Feshbach, Intermediate Structure, \textcolor{blue}{Comments Nucl. Part. Phys. 1, 40 (1967).} 
\bibitem{reaction2} 
H Feshbach, A.K. Kerman, R.H. Lemmer, Intermediate structure and doorway states in nuclear reactions, \textcolor{blue}{Ann. Phys. 41, 230 (1967).} 
\bibitem{reaction3} 
C. Mahaux, Intermediate Structure in Nuclear Reactions, \textcolor{blue}{Ann. Rev. Nucl. Sci. 23, 193 (1973).} 
\bibitem{reaction4} 
A. Mekjian, Nucleon-Nucleus Collisions and Intermediate Structure, \textcolor{blue}{Advances Nucl. Phys. 7, 1 (1973).} 




\end{thebibliography}
\end{document}